\documentstyle [12pt]{article}
\begin{document}

\hfill {WM-98-107}

\hfill {\today}

\vskip 1in   \baselineskip 24pt
{
\Large 
   
   \bigskip
   \centerline{The Anomalous Magnetic Moment of the Muon and}
\centerline{
Higgs-Mediated Flavor Changing Neutral Currents} }
 \vskip .8in
\def\bar{\overline}
 
\centerline{Shuquan Nie and Marc Sher } 
\bigskip
\centerline {\it Nuclear and Particle Theory Group}
\centerline {\it Physics
Department}
\centerline {\it College of William and Mary, Williamsburg, VA
23187, USA}

\vskip 1in
 
{\narrower\narrower In the two-Higgs doublet extension of the
standard model, flavor-changing neutral couplings arise naturally.
In the lepton sector, the largest such coupling is expected to be
$\mu-\tau-\phi$.  We consider the effects of this coupling on the
anomalous magnetic moment of the muon.  The resulting bound on the
coupling, unlike previous bounds, is independent of the value of
other unknown couplings.  It will be significantly improved by the
upcoming E821 experiment at Brookhaven National Lab.}

\newpage

The simplest extension of the standard model involves the addition
of an extra Higgs doublet.   In general, such a model will
generate tree level flavor changing neutral currents (FCNC).  The
Higgs doublet of the standard model does not generate tree level
FCNC because the mass matrix is directly proportional to the
Yukawa coupling matrix, so diagonalization of the former
automatically diagonalizes the latter.   However, in a two-doublet
model, the mass matrix is the sum of two Yukawa coupling matrices
(each times the appropriate vacuum expectation value), and since
the Yukawa coupling matrices are generally not simultaneously
diagonalizable, diagonlization of the mass matrix will {\it not},
in general, diagonalize the Yukawa coupling matrices, leading to
tree-level FCNC.

These tree-level FCNC are dangerous, leading to potentially large
contributions to processes such as $K^o-\bar{K}^o$ mixing. 
This led Glashow and Weinberg\cite{glashowweinberg} to propose a
discrete symmetry, which either couples one Higgs doublet to all
of the fermions (Model I) or else couples one doublet to the
$Q=2/3$ quarks and the other to the $Q=-1/3$ quarks (Model II).

However, it was pointed out by Cheng and Sher\cite{chengsher} 
that, for many models, the coupling was the geometric mean of the
Yukawa couplings of the two fermions.  As a result, FCNC involving
the first generation fields are very small, and the bounds are not
as severe.   Thus, one can also consider Model III, in which no
discrete symmetry is imposed, and the flavor-changing neutral
couplings are simply constrained by experiment.

This has led to a number of calculations involving Model
III [3-11].   The most extensive analyses were
those of Refs.~ \cite{atw3,sheryuan}.   In Ref.
\cite{atw3}, the implications of tree-level FCNC for many
processes were considered, including $K^o-\bar{K}^o$,
$D^o-\bar{D}^o$, $B^o-\bar{B}^o$ and $B^o_s-\bar{B}^o_s$ mixing,
$e^+e^- (\mu^+\mu^-)\rightarrow t\bar{c}+c\bar{t}$,
$Z\rightarrow b\bar{b}$, $t\rightarrow c\gamma$ and the $\rho$
parameter.   In Ref. \cite{sheryuan},  the effects on
$\tau\rightarrow\mu\gamma$, $\tau\rightarrow e\gamma$ and
$\mu\rightarrow e\gamma$, other lepton-flavor violating
decays of the $\tau$ and $\mu$, and a number of rare B-decays were
determined.  In all of these papers, the results were given as
upper bounds on the neutral flavor changing scalar couplings.
In most of these calculations, the results were given in terms of
the product of various Yukawa couplings.  For example, the bound
from $\tau^-\rightarrow \mu^-\mu^+\mu^-$ is dependent on
$h_{\tau\mu}{h_{\mu\mu}}$, where $h_{ij}$ is the coupling of the
scalar to fermions $i$ and $j$.  

In this Brief Report, we point out that a bound on  leptonic
flavor changing couplings can be obtained from the anomalous
magnetic moment of the muon.  This bound has several advantages
over previous bounds:  it depends only on a single Yukawa
coupling, rather than a product; it is stronger than previous
bounds, using reasonable assumptions about the ratio of couplings;
it could be improved significantly in the near future at the
upcoming experiment E821\cite{expe} at Brookhaven National Lab.

As shown in the above references, one can choose a basis for the
two Higgs doublets such that only one doublet, $H$, obtains a vacuum
expectation value.  That doublet will then have flavor-diagonal
couplings, while the couplings of the other, $\phi$, will be
unconstrained.    The
relevant term in the Lagrangian is
\begin{equation}
m_e\bar{e}_Le_R(\sqrt{2}H/v)+
m_\mu\bar{\mu}_L\mu_R(\sqrt{2}H/v)+
m_\tau\bar{\tau}_L\tau_R(\sqrt{2}H/v)+
h_{ij}\bar{l}_{iL}l_{jR}\phi + {\rm H.c.}
\end{equation}
The
$\phi$ field is composed of a scalar
$\phi_s$ and a pseudoscalar $\phi_p$.
 Since one expects\cite{chengsher} the
heavier generations to have larger flavor-changing couplings, we
will look at the bound on the $h_{\mu\tau}$ arising from the
magnetic moment of the muon.

The diagram is given in Figure 1.   The diagrams in which the
photon couples to the external lines do not give a contribution
to the magnetic moment of the muon.  The calculation is
straightforward, and we find that the contribution to $a_\mu\equiv
{g_\mu-2\over 2}$ is given by
\begin{equation}
a_{\mu}={h_{\mu\tau}^2\over 16\pi^2}\int_0^1
{z^2(1-z) \pm  z^2{m_\tau\over m_\mu} \over
z(z-1)+z{m_\tau^2\over m_\mu^2}+(1-z){m^2_\phi\over m^2_\mu}}\ dz
\end{equation}
where $m_\phi$ is the mass of the scalar or pseudoscalar, and the
$+(-)$ sign is chosen for the scalar (pseudoscalar).

The resulting contribution is given in Figure 2.  The scalar and
pseudoscalar have contributions which are almost identical
(within a few percent), but of opposite sign; the scalar
contribution is shown.  Since there is no reason that the masses
should be similar, and since the result is so sharply dependent on
the mass, we expect the lighter of the two to make the dominant
contribution.  

Consider the case in which the lighter scalar is $80$ GeV
(current LEP bounds apply to a standard model Higgs, and are
generally weaker for two doublet models).  We see that this gives a
value of
$a_\mu$ which is
$1.14\times 10^{-6} h_{\mu\tau}^2$.  The current value of $a_\mu$ is
in agreement with expectations, and the uncertainty is
$8.4\times 10^{-9}$.  Thus, the contribution of the flavor-changing
coupling must be less than this uncertainty, or $h_{\mu\tau}$ must
be less than $0.12$. More importantly, the experimental uncertainty
in the anomalous magnetic moment will shortly decrease by up to a
factor of 20, so that a bound on $h_{\mu\tau}$ of $0.027$ can be
obtained.

How does this bound compare with other bounds?  As noted above, 
{\bf all} other bounds depend on the product of $h_{\mu\tau}$
times other unknown Yukawa couplings.  Thus, the bound is unique. 
However,  one can make reasonable assumptions about these other
couplings.  For example, in Ref. \cite{sheryuan}, it was argued
that grand unified theories will give a relationship between
$h_{bs}$ and $h_{\mu\tau}$; from this, they looked at the process
$B\rightarrow K\mu\tau$ to get the bound $h_{\mu\tau} < 0.024$. 
Unfortunately, there are no experimental limits listed for
$B\rightarrow K\mu\tau$; their bound  was obtained by noting that
$17\%$ of $\tau$'s decay into $\mu$'s, and using the bound on
$B\rightarrow \mu^+\mu^-X$.  Since the experimental cuts would be
quite different, this result is very uncertain, and thus the
bound on $h_{\mu\tau}$ could be considerably higher (and also
requires the assumption of grand unification).  In addition, one can
assume that the ratio  of
$h_{\tau\tau}$ to
$h_{\mu\tau}$ is $\sqrt{m_\tau\over m_\mu}$, in which case the
bound on $h_{\mu\tau}$ from $\tau \rightarrow 3\mu$ gives $0.2$,
which is also much weaker\cite{caveat}.   It should be pointed out
that a similar diagram could bound the flavor-changing $h_{\mu e}$
coupling, but much stronger bounds on that can be obtained from
bounds on radiative muon decay.

In the absence of any solid theoretical understandings of 
Yukawa couplings, one must rely on experiment to bound such
couplings.  In the simplest extension of the standard model, a
flavor changing coupling of the $\mu$ and $\tau$ to a neutral
scalar can, in general, exist.  In this work, we have shown that
the strongest bound on such a coupling (independent of assumptions
about other couplings) arises from the contribution to the
anomalous magnetic moment of the muon.  Should the E821 experiment
find a value which is in conflict with the standard model, this
may provide one of the simpler explanations.

We thank Lee Roberts for getting us interested in the anomalous
magnetic moment, and Chris Carone for useful discussions.  This
work was supported by the National Science Foundation.

 \def\prd#1#2#3{{\rm Phys. ~Rev. ~}{\bf D#1}, #3 (19#2)}
\def\plb#1#2#3{{\rm Phys. ~Lett. ~}{\bf B#1}, #3 (19#2) }
\def\npb#1#2#3{{\rm Nucl. ~Phys. ~}{\bf B#1}, #3 (19#2) }
\def\prl#1#2#3{{\rm Phys. ~Rev. ~Lett. ~}{\bf #1}, #3 (19#2) }

\bibliographystyle{unsrt}

\begin{thebibliography}{99}
\bibitem{caveat}
It was argued by Cheng and
Sher\cite{chengsher} that the most natural value for
$h_{\mu\tau}$ is the geometric mean of the Yukawa couplings of the
muon and tau, or $0.0025$.  Our bound is still above that value. 
However, the symmetry arguments in that work apply only to the
ratio of the couplings, not their overall scale, and thus higher
couplings are certainly possible.


\newpage
\end{thebibliography}

\newpage

\begin{figure}
\caption{Contribution to the anomalous magnetic moment of the muon
from the exchange of flavor-changing scalars.  The scalar can be
either a scalar or a pseudoscalar.}
\end{figure}

\begin{figure}
\caption{The contribution of the diagram of Figure 1 to the
anomalous magnetic moment of the muon.  The contribution of the
scalar is shown; that of the pseudoscalar is within a few percent
of that of the scalar, but of opposite sign.}
\end{figure}

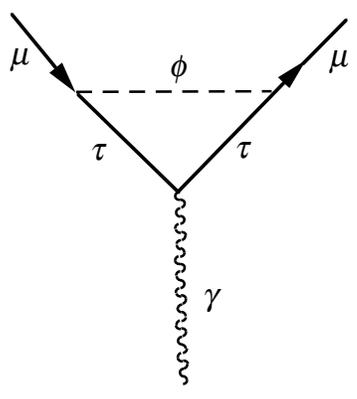
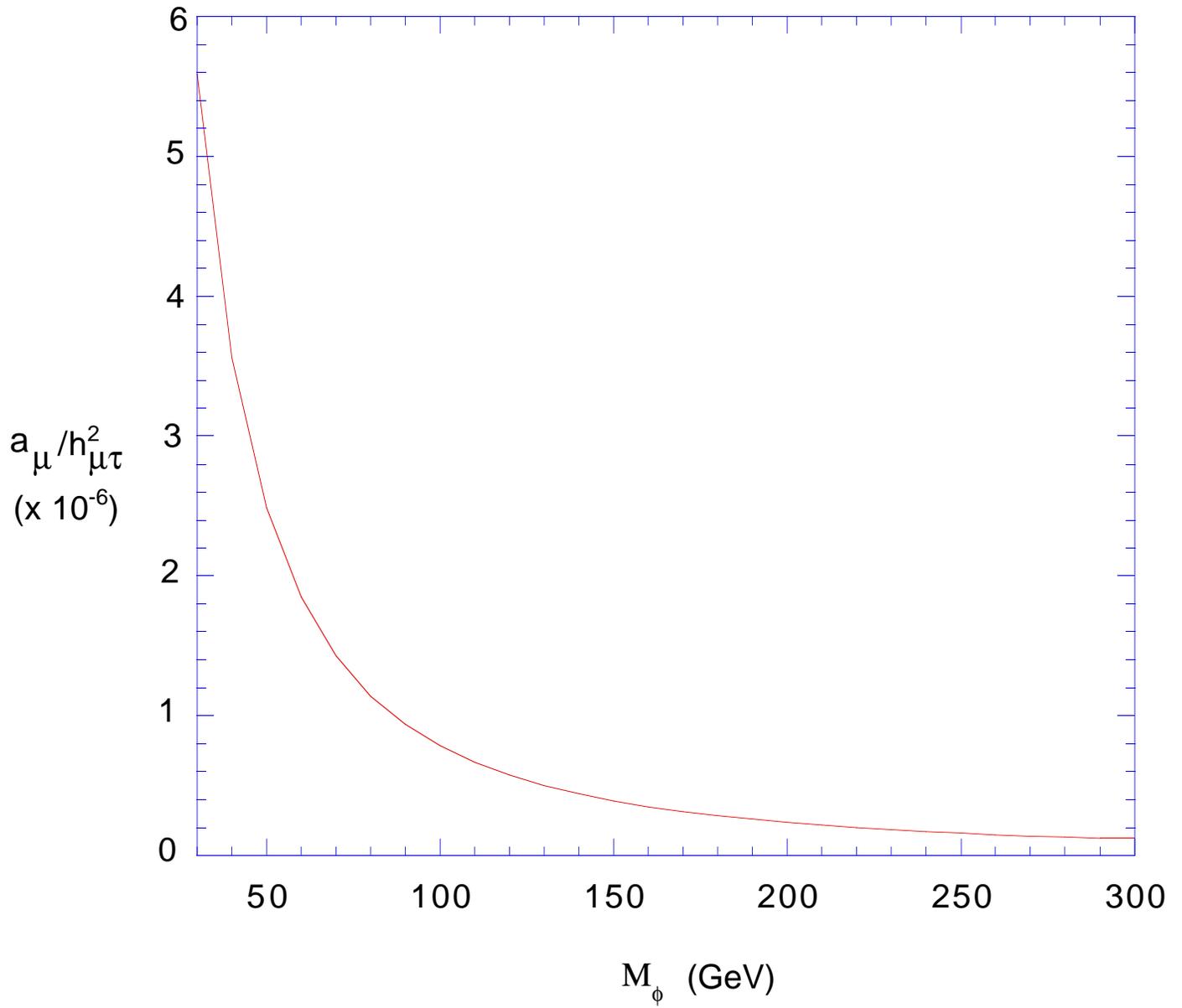
\end{document}